\documentclass[11pt]{article}

\usepackage[margin=1in]{geometry}
\usepackage{enumitem}
\usepackage{bm}
\usepackage{multirow}
\usepackage{graphicx}
\usepackage[boxed]{algorithm}

\usepackage{mathptmx}
\usepackage{latexsym}
\usepackage{amsmath,amsfonts,amsthm,amssymb}
\usepackage{verbatim}
\usepackage{xspace}
\usepackage[dvipsnames,usenames]{color}

\def \sumds {\displaystyle\sum}

\newtheorem{theorem}{Theorem}
\newtheorem{lemma}[theorem]{Lemma}
\newtheorem{corollary}[theorem]{Corollary}

\newtheorem{observation}[theorem]{Observation}

\newcommand{\gma}{{\sc umsc}\xspace}
\newcommand{\ocg}{{\sc ocg}\xspace}
\newcommand{\ompc}{{\sc ompc}\xspace}
\newcommand{\ocl}{{\sc ocl}\xspace}
\newcommand{\oce}{{\sc oce}\xspace}

\newcommand{\simload}{\tilde{L}}

\newcommand{\eat}[1]{}

\newcommand{\opt}{{\text{\sc opt}}}
\newcommand{\pr}{{\mathbb{P}}}
\newcommand{\ex}{{\mathbb{E}}}

\renewcommand{\vec}[1]{{\bf #1}}

\title{Online Covering with Convex Objectives and Applications}

\author{
	Yossi Azar\thanks{Email: azar@tau.ac.il. Supported in part by the Israel Science
		Foundation (grant No. 1404/10) and by the Israeli Centers of Research Excellence (I-CORE) program, (Center  No.4/11). 
		Part of this work was	done while the author was visiting Microsoft Research, Redmond.}
\\
	Tel-Aviv University \\
	\and
	Ilan Reuven Cohen\thanks{Email: ilanrcohen@gmail.com. Supported in part by the 
	Israeli Centers of Research Excellence (I-CORE) program.}\\
Tel-Aviv University \\
		\and
		Debmalya Panigrahi\thanks{Email: debmalya@cs.duke.edu. Supported in part by 
		a Duke University startup grant and a Google Faculty Research Award.
		Part of this work was	done while the author was visiting Microsoft Research, Redmond.}\\
		Duke University \\
}

\date{}

\begin{document}

\maketitle

\begin{abstract}
We give an algorithmic framework for minimizing 
general convex objectives (that are differentiable and 
monotone non-decreasing) over a set of covering constraints
that arrive online. This substantially extends previous 
work on online covering for linear objectives 
(Alon {\em et al.}, STOC 2003)
and online covering with offline packing constraints
(Azar {\em et al.}, SODA 2013). 
To the best of our knowledge, this is the first
result in online optimization for 
generic non-linear objectives; special cases
of such objectives have previously been considered,
particularly for energy minimization.

As a specific problem in this genre, we 
consider the unrelated machine scheduling problem 
with startup costs and arbitrary $\ell_p$ norms
on machine loads (including the surprisingly 
non-trivial $\ell_1$ norm representing total
machine load). This problem was studied earlier
for the makespan norm in both the offline
(Khuller~{\em et al.}, SODA 2010; Li and 
Khuller, SODA 2011) and online settings
(Azar {\em et al.}, SODA 2013). We adapt 
the two-phase approach of obtaining a fractional
solution and then rounding it online (used 
successfully to many linear objectives) to
the non-linear objective.  The fractional 
algorithm uses ideas from our general 
framework that we described above (but does not fit the framework
exactly because of non-positive entries
in the constraint matrix). The rounding 
algorithm uses ideas from 
offline rounding of LPs with non-linear objectives
(Azar and Epstein, STOC 2005; 
Kumar {\em et al.}, FOCS 2005). Our competitive
ratio is tight up to a logarithmic factor. 
Finally, for the 
important special case of total load 
($\ell_1$ norm), we give a different rounding 
algorithm that obtains a better competitive ratio
than the generic rounding algorithm for $\ell_p$ 
norms. We show that this competitive ratio is 
asymptotically tight.  
\end{abstract}

\thispagestyle{empty}
\clearpage
\setcounter{page}{1}
\clearpage

\section{Introduction}
\label{sec:intro}

Positive linear programming (also known as packing/covering) 
with convex (non-linear) objectives model a wide range of problems 
in combinatorial optimization and operations research. In algorithmic 
theory, they have been used in many areas including 
machine scheduling~\cite{AzarE05}, 
packet routing~\cite{AntoniadisIKMNPS14}, 
energy minimization~\cite{BansalGKNPS12}, etc. 
In this paper, we consider the problem of minimizing arbitrary convex 
functions under linear covering constraints that arrive online. This significantly
generalizes and extends previous frameworks for online covering algorithms 
with linear objectives~\cite{AlonAABN09,BuchbinderN09} and with offline packing 
constraints~\cite{AzarBFP13}.
For convex objectives that are monotone and differentiable, 
we give a simple deterministic online algorithm that guarantees a nearly 
optimal solution.
Then, we consider a natural representative of this 
genre of problems in machine scheduling --- minimize the $\ell_p$ norm of 
machine loads where each machine has a startup cost. This problem arises in 
the context of energy optimization in cloud computing, and was previously studied
for the makespan norm of machine loads in both the offline~\cite{KhullerLS10,LiK11} 
and online~\cite{AzarBFP13} settings.
We give an online algorithm for this problem based on a two-phase process
(commonly used in the online setting for {\em linear} objectives) of 
obtaining a competitive fractional solution, and rounding it online.
While our online framework for general convex objectives cannot be used 
directly,\footnote{Some of 
	the constraints are not packing/covering constraints, i.e., have negative
	coefficients.} we use the intuition that we gained from it 
to obtain an online fractional solution in the first phase. 
In the second phase, we combine ideas from offline 
rounding for $\ell_p$ objectives~\cite{AzarE05,KumarMPS05} and online rounding for 
exponential objectives~\cite{AzarBFP13} in a novel manner to obtain an 
integral assignment of jobs to machines.

\medskip
\noindent
\textbf{Online Covering with General objectives}~(\ocg):
The goal is to minimize a {\em convex}, {\em non-decreasing}, 
{\em differentiable} function $f(\vec{x})$ of $m$ variables 
$\vec{x} = \langle x_1, x_2, \ldots, x_m\rangle$ subject to 
$n$ linear covering constraints $C\vec{x} \geq \vec{c}$
that arrive online.
Here, $C$ is an $n \times m$ matrix and $\vec{c}$ is an 
$n$-dimensional vector, both with non-negative entries.
The variables $x_i$, $1\leq i\leq m$, are also constrained 
to be non-negative and must be monotonically non-decreasing 
over the course of the online algorithm. 
On the arrival of a new covering constraint, it must be 
satisfied by increasing the values of the variables
(note that the monotonicity of the variables and non-negativity
of the constraint matrix implies that all constraints previously
satisfied continue to be satisfied). 
This framework generalizes the following settings:
\begin{itemize}
	\item \textbf{Online Covering with Linear Objectives}~(\ocl) \cite{BuchbinderN09,AlonAABN09}:
	This is the special case where the function $f(\vec{x})$ 
	is a linear function. This problem, in turn, generalizes
	the fractional versions of several important problems such
	as online set cover~\cite{AlonAABN09}, online non-metric
	facility location~\cite{AlonAABN06}, online network design 
	problems~\cite{AlonAABN06,NaorPS11,HajiaghayiLP13,HajiaghayiLP14},
	etc.
	\item \textbf{Online Mixed Packing and Covering}~(\ompc) \cite{AzarBFP13}:
	In this problem, there are two sets of constraints: a set of $n$ linear
	covering constraints $C \vec{x} \geq \vec{c}$ that arrive online and 
	a set of $r$ linear packing constraints $P\vec{x} \leq \vec{p}$ 
	that are given offline. All entries in $C, P, \vec{c}$, and $\vec{p}$
	are non-negative, and the variables $x_i$, $1\leq i\leq m$, must
	be non-negative and monotonically non-decreasing over the course
	of the online algorithm. The goal here is to exactly satisfy all
	the covering constraints, and approximately satisfy all the packing
	constraints (the approximation is provably required). For convenience,
	let us define a new set of (derived) variables 
	$\vec{\lambda} = \langle \lambda_1, \lambda_2, \ldots, \lambda_r \rangle$,
	where $\lambda_k = \frac{\sum_i P_{ki} x_i}{p_k}$. In other 
	words, $\lambda_k$ is the {\em violation}\footnote[2]{One may also define 
	$\lambda_k = \max\left(\frac{\sum_i P_{ki} x_i}{p_k}, 1\right)$; 
	our results hold also for this definition.} for the $k$th packing 
	constraint. Then, the objective is to minimize the maximum violation;
	i.e., $f(\vec{x}) = \max_k \lambda_k$. This objective, as stated, 
	has a large ($O(r)$) measure of convexity (will be defined later)
	and hence it is not useful for the \ocg framework. However, 
	as shown in \cite{AzarBFP13}, 
	the objective function can be modified to  
	$f(\vec{x}) = \ln \left(\sum_k e^{\lambda_k} \right)$ up to a 
	loss of $O(\log r)$ in the competitive ratio. The new function satisfies
	the conditions of the \ocg problem. More generally, we can also consider 
	any $\ell_p$	norm of the vector $\vec{\lambda}$ as our function; this 
	also generalizes \cite{AzarBFP13} since the maximum violation is known 
	to be within a constant factor of the $\ell_{\ln r}$ norm.
\end{itemize}	

The second part of our paper focuses on a representative problem in 
the genre of online covering problems with non-linear objectives:
\textbf{Unrelated Machine Scheduling with Startup Cost}~(\gma).
Let $M$ be a set of $m$ machines, where machine $i$ has {\em startup cost}
$c_i\geq 0$, and $J$ be a set of $n$ jobs that arrive online.
The {\em processing time} of job $j$ on machine $i$ is denoted $p_{ij} \geq 0$.
A {\em schedule} is an assignment of jobs to machines, and the 
{\em load} $L_i$ of a machine
$i$ in a schedule is the sum of processing times of all jobs assigned to it.
The {\em open machines} $M_o$ are the machines to which at least one job has
been assigned
and the cost of the schedule is the sum of startup costs of open machines.
The goal is to obtain a schedule that simultaneously minimizes cost
and some function $f$ of machine loads. The typical functions for $f$ are:
(1) the {\em makespan} or  maximum load among all machines, i.e.,
$f = \max_{i\in M} L_i$, 
(2) the total load over all machines, i.e.,
$f = \sum_{i\in M} L_i$, and 
(3) the more general $\ell_p$-norm of the load, i.e.,
$f = (\sum_{i\in M} (L_i)^p)^{1/p}$ for any fixed $p \in [1, \log m]$ 
(since $\ell_p \equiv \ell_{\infty}$ for $p \geq \log m$).
The existence of startup costs makes 
even the case of minimizing the total load ($\ell_1$ norm) 
non-trivial since the machine on which a job 
runs the fastest might have a large cost. This forces 
the algorithm to strike a balance
between opening machines that have large startup costs but can
run jobs at high speeds and those that have smaller startup
costs but run slower.

\noindent
{\bf Note:} We assume that the \gma input includes
a pair of values $({\bf C}, {\bf L})$ with the guarantee that
there exists a schedule of cost at most $\bf C$ and
$\ell_p$-norm at most $\bf L$ ($p$ is fixed).
Using standard doubling guesses,
our formulation can be shown to be aymptotically equivalent 
to one where 
one objective needs to be
optimized subject to a given bound on the other. Moreover,
our formulation subsumes single objective formulations where
the two objectives are combined using a linear
function.

The \gma problem is closely connected to energy management in
data centers, which has recently emerged as one of the 
most important practical challenges in cloud computing 
(see, e.g.,~\cite{BirmanCR09} for a discussion). 
With this motivation, the problem was
studied in the offline setting for the 
makespan norm~\cite{KhullerLS10,Fleischer10,LiK11} 
and in the online setting~\cite{AzarBFP13}.
In this paper, we extend this line of work significantly 
to all $\ell_p$ norms, including the surprisingly non-trivial
$\ell_1$ norm representing total machine loads.
Note that the \gma problem generalizes the online set 
cover problem~\cite{AlonAABN09} (for $p_{ij} = 0$ or $\infty$) 
and the online unrelated machine scheduling 
problem~\cite{AspnesAFPW97,AwerbuchAGKKV95}
(for $c_i = 0$). A similar energy minimization problem 
(call it {\em online covering for energy minimization}
or \oce)
was studied in \cite{GuptaKP12} which can be thought of
as the \gma problem with assignment costs instead of startup
costs. This seemingly minor difference, however, completely 
changes the structure of the problem since the \oce problem
does {\em not} generalize set cover. In fact, perhaps the 
most illustrative difference between the two problems 
is for the case of linear loads, where \gma remains non-trivial
whereas a greedy algorithm suffices for \oce. Moreover, the goal in 
\cite{GuptaKP12} was to only obtain a fractional solution
whereas we are interested in an integral solution and therefore
need to consider integrality gaps.

\subsection{Our Results}

\textbf{The \ocg framework.}
We will denote the maximum and minimum non-zero entry in 
the constraint matrix $C$ by $c_{\max}$ and $c_{\min}$
respectively. Our result also depends on two parameters
of the objective function $f$. The first parameter 
$\beta = \displaystyle\max_{\vec{x}} \frac{\sum_{i=1}^m x_i \cdot \frac{\partial f}{\partial x_i}}{f(\vec{x})}$. 
Informally, this is a measure of the convexity of the function: e.g.,
$\beta = O(1)$ for any polynomial function but infinite for exponential
functions. The second parameter $\gamma$ is the smallest positive 
number such that $f(1/\gamma, \dots, 1/\gamma) \leq \opt$. 
To understand the dependence on $\gamma$, consider an 
objective $f(x_1, x_2) = 0$ if $x_1 = 0$ or $x_2 = 0$
but $> 0$ otherwise. For this objective, it is impossible
to obtain a finite competitive 
ratio\footnote[1]{To see this, let the first constraint be
$x_1 + x_2 \geq 1$. 
If the online algorithm sets $x_1 > 0$ (resp., $x_2 > 0$)
in response to this constraint, then the next constraint
is $x_2 \geq 1$ (resp., $x_1 \geq 1$).} and this is 
encapsulated by an infinite value of $\gamma$.

We are now ready to state our result.
\begin{theorem}
	\label{thm:mainocg}
	There is a deterministic online algorithm for the \ocg problem
	that produces a fractional solution with objective at most 
	$f(\beta \log(\gamma/c_{\min}) \vec{x^*}) + \beta f(\vec{x^*})$, 
	where $x^*$ is any optimal solution. 
	In particular, for sub-homogeneous functions (i.e., 
	functions satisfying 	$f(\eta \vec{x}) \leq \eta f(\vec{x})$ 
	for any $\eta > 1$) the competitive ratio is 
	$O(\beta \log (\gamma/c_{\min}))$.
\end{theorem}
\noindent
First, we apply Theorem~\ref{thm:mainocg} to a linear 
objective $\sum_{i=1} a_i x_i$, i.e., 
the \ocl problem~\cite{BuchbinderN09}.
We note that any variable $x_i$ for which 
$a_i/c_{\max} > \opt$ can be discarded at the outset.
After discarding these variables, we can set 
$\gamma = m c_{\max}$, and the competitive ratio is 
$O(\log (m c_{\max} / c_{\min}))$ since $\beta = 1$.
For $\{0, 1\}$ constraint matrices (e.g. the fractional
set cover problem~\cite{AlonAABN09} and network design 
problems in \cite{AlonAABN06,NaorPS11,HajiaghayiLP13,HajiaghayiLP14}), 
the competitive ratio is $O(\log m)$.

Next, we consider the \ompc problem with the $\ell_p$ 
norm objective, i.e., 
$f(\vec{x}) = \left(\sum_k (\lambda_k)^p\right)^{1/p}$ 
(recall that $\lambda_k$ denotes the violation of the 
$k$th packing constraint). 
Let $p_{\max}$ and $p_{\min}$ be the 
maximum and minimum non-zero entries in the packing
matrix $P$ respectively, and 
let $\kappa = p_{\max}/p_{\min}$; similarly, let
$\rho = c_{\max}/c_{\min}$. Also, let
$d \leq m$ denote the maximum number of variables in any 
packing or covering constraint.
In order to apply Theorem~\ref{thm:mainocg}, we set 
$\gamma = d \cdot c_{\max} \cdot (p_{\max}/p_{\min})$,
since for any packing constraint $k$, we have 
$\sum_{i=1}^m p_{ki} x^0_i \leq p_{\min}/c_{\max} \leq \sum_{i=1}^m p_{ki} x^*_i$. 
Also, $\beta = p$ for the $\ell_p$  norm function, 
yielding the following corollary (note that, it is enough to consider
$p \leq \log r$ since $\ell_p \approx \ell_{\log r}$ for any $p \geq \log r$, 
including the $\ell_\infty$ norm).
\begin{corollary}
\label{cor:lp}
	There is a deterministic online algorithm for the \ompc problem
	with $\ell_p$ norm that has a competitive ratio of
	$O(p\log (d\rho\kappa))$. For $\{0,1\}$ constraint matrices, 
	the competitive ratio is $O(p\log d)$.
\end{corollary}
\noindent
This matches the upper bound of $O(\log r\cdot \log (d\rho\kappa))$ 
for the $\ell_{\infty}$ norm
($\max_k \lambda_k$) in \cite{AzarBFP13} 
by using $p = \log r$ since the $\ell_{\log r}$ norm approximates the 
$\ell_\infty$ norm up to a small constant.
Alternatively, one may try to apply Theorem~\ref{thm:mainocg} directly 
for the function $f(\vec{x}) = \max_k \lambda_k$. 
However, this results in a worse approximation ratio since for 
this function, $\beta = r$. In fact, the authors in \cite{AzarBFP13} 
used a third function $f(\vec{x}) = \ln \left(\sum_k e^{\lambda_k}\right)$
as the surrogate objective for $\max_k \lambda_k$. Theorem~\ref{thm:mainocg}
can be directly applied to this function as well, yielding a matching
result to those obtained by the $\ell_{\log r}$ norm in 
Corollary~\ref{cor:lp} and in \cite{AzarBFP13}.

We also show that Corollary~\ref{cor:lp} is nearly tight, by adapting a
lower bound in \cite{AzarBFP13} to the $\ell_p$ norm.
\begin{theorem}
	\label{thm:mainlowerompc}
	Any deterministic algorithm for \ompc with respect to the $\ell_p$ norm on $\lambda$ is
	$\Omega(p \log(d/ \log r) )$-competitive for $p\leq \log r$, even for $\{0, 1\}$
	constraint matrices.
\end{theorem}

\medskip
\noindent
\textbf{The \gma problem.} 
Following standard convention, we say that a randomized
algorithm for the \gma problem has a bi-criteria competitive
ratio of $(\alpha, \beta)$ if it produces a schedule of
expected cost at most $\alpha {\bf C}$ and
the expected $\ell_p$ norm of the load is 
at most $\beta {\bf L}$. Our main result
is a randomized algorithm that proves the next theorem.
\begin{theorem}
	\label{thm:main}
	There is a randomized online algorithm for the \gma problem
	for arbitrary fixed $p$ with a competitive ratio of
	$(O(\log m \log (mn)), O(p^2 \log^{1/p} (mn)))$.
\end{theorem}
\noindent
Since $p\leq \log m$, our competitive ratio
is upper bounded by  
$(O(\log m \log (mn)), O(\log^2 m\log^{1/p} (mn))$.

Recall that the \gma problem generalizes the set cover problem~\cite{AlonAABN09}
and the unrelated machine scheduling problem for $\ell_p$ 
norms~\cite{AwerbuchAGKKV95, Caragiannis08}.
The lower bound for the \gma problem is
derived from lower bounds for these problems (see
\cite{AlonAABN09,Korman05} for the 
cost lower bound derived from online set cover and
\cite{AwerbuchAGKKV95, Caragiannis08} for the $\ell_p$-norm lower
bound derived from online unrelated machine scheduling).
\begin{observation}
	\label{thm:lowerbound}
	No algorithm for the \gma~problem can have a competitive
	ratio of $o(p)$ in the $\ell_p$-norm of machine loads.
	Further, under standard complexity assumptions, no algorithm
	for this problem can have a competitive ratio of $o(\log m \log n)$
	in the cost of the schedule.
\end{observation}
\noindent
It follows from these lower bounds that the competitive ratios in
Theorem~\ref{thm:main} are almost tight in both objectives.

We also separately consider the important special case of $p=1$,
where the goal is to minimize the sum of all machines loads.
For this case, Theorem~\ref{thm:main} gives a competitive
ratio of $(O(\log m \log (mn)), O(\log (mn)))$. We improve this
result and obtain a tight (up to constants) competitive ratio in
both objectives.
\begin{theorem}
	\label{thm:l1}
	There is a randomized online algorithm for the \gma problem for
	$p=1$ with a competitive ratio of
	$(O(\log m\log n), O(1))$.
\end{theorem}

\subsection{Our Techniques}

To solve the \ocg problem, we use a continuous algorithm where the values 
smoothly increase over time. (The algorithm can be discretized for polynomial
implementation, but the continuous version is easier to describe.) 
The rate of increase of each variable is inversely proportional the current 
partial derivative of the objective for this variable. Note that this extends 
the algorithm for online set cover~\cite{AlonAABN09} 
where the partial derivative is the cost 
of the set. In the analysis,  we implicitly use the Lagrangian dual of the 
convex objective. The algorithm increases the dual variable of the current 
constraint at unit rate (as in \cite{AlonAABN09,BuchbinderN09}). 
The analysis establishes approximate stationarity of the optimal solution, and 
a relationship between the growth of the primal objective and the Lagrangian
dual. These two facts are coupled to bound the value of the objective in the 
algorithmic solution by that of any suitably scaled feasible solution, thereby
showing Theorem~\ref{thm:mainocg}. 
    
For the \gma problem, 
using the syntactic definition of the $\ell_p$-norm
(we actually use the $\ell_p^p$ norm for ease of 
manipulation)
as the objective function leads to a polynomial integrality gap.
Consider the following simple example.
Suppose there are $m$ machines with startup cost $1$ each and
$m$ jobs arrive with $p_{ij} = 1$ for each $(i, j)$-pair. Also,
let ${\bf C} = 1$. Then, a feasible fractional solution is to
open each machine to $x_i = 1/m$ and set $y_{ij} = 1/m$ for
each $(i, j)$-pair. While the objective value of this fractional
solution is $m$, any integer solution with a poly-logarithmic
competitive ratio in the cost (recall that this is what we are
aiming for) can open at most a poly-logarithmic number of machines,
and therefore will have an objective value of at least
$\log m (m/log m)^p$. To overcome this integrality gap, we refine
our definition of the $\ell_p$-norm of the load on a partially
open machine (for a fully open machine, we continue to use the
syntactic definition of
$\sum_{i\in M}\left(\sum_{j\in J} p_{ij} y_{ij}\right)^p$)
to $\sum_{i\in M}\left(\frac{\sum_{j\in J} p_{ij} y_{ij}}{x_i}\right)^p x_i$,
where $y_{ij}$ is the assignment of job $j$ to machine $i$
and $x_i$ is the fraction to which machine $i$ is open.
(We use constraints $y_{ij} \leq x_i$ which deviates from positive LPs
as stated above.)

However, there is still a large integrality gap
since a fractional solution can split a large job into
several small jobs and distribute them on multiple machines.
In order to overcome it, we add an extra term
$\sum_{i\in M}\left(\sum_{j\in J} y_{ij} p_{ij}^p\right)$
to the objective function (see also ~\cite{KumarMPS05,AzarE05}). 
Note that for an integer solution, this
additional term is bounded above by the actual $\ell_p^p$ norm.
The complete LP is given in Fig.~\ref{fig:lp}.

\begin{figure}[!htb]
		\centering
		\small
		Minimize $\quad \sum_{i\in M} \left(\frac{\sum_{j\in J} p_{ij} y_{ij}}{x_i}\right)^p x_i + \sum_{i\in M}\left(\sum_{j\in J} y_{ij} p_{ij}^p\right) \quad$ subject to
		\begin{eqnarray}
		\label{eqn:cost}      \sum_{i\in M} c_i x_i  & \le & {\bf C} \\
		\label{eqn:fraction}		y_{ij} & \leq & x_i\quad \forall~i\in M,~j\in J \\
		\label{eqn:covering}		\sum_{i\in M} y_{ij} & \geq & 1\quad \forall~j\in J \\
		\label{eqn:range1}		
		x_i, y_{ij} & \in & [0, 1] 
		\end{eqnarray}
		\caption{\small The \gma LP}
		\label{fig:lp}
	\end{figure}

To obtain a fractional solution for this formulation, we design a non-linear
potential function that guides multiplicative updates of the variables.
For partially open machines, the potential function
is defined according to the fractional cost of the machine; during
this phase, the primary
goal of the algorithm is cost minimization. The multiplicative
update steps are designed such that the load on the machine is
``small'' in this phase.
Once $x_i$ increases to 1, i.e., machine $i$ is {\em fully open},
the potential function is defined on the $\ell_p$-norm of the
fractional load on the machine. In this phase, the primary goal of
the algorithm shifts to load minimization.
In bounding the $\ell_p$-norm
of the load, we also use ideas due to
Caragiannis~\cite{Caragiannis08}, who gave an elegant analysis
for the problem without startup costs.

\subsection{Previous Work}

Packing and covering have been widely used and analyzed in offline 
scenarios, typically for linear objectives 
(e.g.~\cite{Plotkin95fastapproximation,GargK98}).
In a sequence of recent papers, online versions of these problems 
have also been studied including
online set cover~\cite{AlonAABN06}, 
network design~\cite{AlonAABN09,NaorPS11,HajiaghayiLP13,HajiaghayiLP14},
paging~\cite{BansalBN12a,BansalBN12b,BansalBN10,BansalBMN11}, 
general online covering  {\em or} online packing constraints~\cite{BuchbinderN09},
online covering constraints {\em and} offline packing constraints~\cite{AzarBFP13},
etc. Non-linear objectives have also been considered for specific problems, 
especially related to energy minimization (e.g., \cite{GuptaKP12}).
To the best of our knowledge, this is the first paper to give results
for optimizing general non-linear objectives under linear constraints.

Assigning jobs that arrive online to unrelated machines
so as to minimize the $\ell_p$-norm of machine loads is a
central question in scheduling theory. For $p=1$, the natural
greedy strategy of assigning each job to the machine on which
it runs the fastest is optimal, but for $p > 1$, the problem
turns out to be more challenging. 
For the makespan objective (maximum load or the 
$\ell_{\infty}$ norm, which is also asymptotically equivalent
to any $\ell_p$ norm with $p\geq \log m$), 
Aspnes~{\em et al.} \cite{AspnesAFPW97} obtained a
competitive ratio of $O(\log m)$,
which is asymptotically tight~\cite{AzarNR95}. 
For any $p\leq \log m$,
Awerbuch~{\em et al}~\cite{AwerbuchAGKKV95} obtained
a tight competitive ratio of $O(p)$.
Subsequently,
Caragiannis~\cite{Caragiannis08,CaragiannisFKKM06} provided
an elementary analysis for this algorithm, while also
tightening the constants in the upper and lower bounds. 
Various other models and objectives have been considered 
for the load balancing
problem; the interested reader is referred to surveys
such as \cite{Azar96,Sgall96,PruhsST04,Sgall05}.

The offline version of \gma with the makespan objective was introduced by 
Khuller~{\em et al.} \cite{KhullerLS10}, where they gave an
$O(2(1+1/\epsilon)(1 + \ln (n/OPT)), 2+\epsilon)$-approximation algorithm for
any $\epsilon > 0$. (For further work on this problem, see
\cite{Fleischer10,LiK11}). The online version of this problem
with the makespan objective was considered in \cite{AzarBFP13}, who obtained
a poly-logarithmic bicriteria competitive ratio.
We significantly generalize these results by considering $\ell_p$-norms for 
arbitrary values of $p$.

\medskip
\noindent
{\bf Roadmap.} The algorithm for \ocg (Theorem~\ref{thm:mainocg}) is in
Section~\ref{sec:ompc}. The fractional algorithm and the 
randomized rounding procedure for \gma with general $\ell_p$ norms
(Theorem~\ref{thm:main}) are 
in Sections~\ref{sec:fractional} and \ref{sec:rounding} respectively. 
The lower bound for \ompc (Theorem~\ref{thm:mainlowerompc})
is given in the appendix.

\section{Algorithm for the \ocg problem}
\label{sec:ompc}

We consider the convex program for an $m$-dimensional 
{\em non-negative}
variable $\vec{x} = \langle x_i: 1 \leq i \leq m\rangle$: 
\begin{center}
minimize $f(\vec{x})$ subject to
	$C \vec{x} \geq \vec{1}$,
\end{center}
where 
	the objective function $f$ is {\em convex}, {\em monotone 
	non-decreasing}, and {\em differentiable everywhere}.
	The covering matrix $C$ is an $m \times n$-dimensional 
	non-negative matrix (the $(i, j)$th entry is denoted $c_{ij}$) and
	the RHS is wlog (by scaling) the all-ones vector in 
	$n$ dimensions.
The constraints arrive online and must be satisfied when
they arrive. The variable $\bf x$ has to be monotone 
non-decreasing over time in every dimension.
It will also be convenient to define the Lagrangian dual:
\begin{equation*}
	L(\vec{x}, \vec{y}) = f(\vec{x}) - \vec{y} \cdot (C \vec{x} - \vec{1}).
\end{equation*}

\subsection{Description of the Algorithm} 
We define a continuous algorithm where $\bf x$
is initialized to a certain value and smoothly increases 
over time. For a polynomial implementation, this 
algorithm can be discretized by choosing a small enough 
discrete ``step size''. 

We initialize $\bf x$ to the vector $\vec{x^0} = (1/\gamma,1/\gamma,\dots ,1/\gamma)$,
where $\gamma$ is large enough so that 
\begin{equation*}
	f(1/\gamma,1/\gamma,\dots ,1/\gamma) \leq \opt.
\end{equation*}	
%
%
When a constraint $\sum_i c_{ij} x_i \geq 1$ arrives online, 
we increase $\vec{x}$ at the following rate until the 
constraint is satisfied:
\begin{equation*}
	\forall~i\in [m], ~\frac{d x_i}{dt} = \frac{c_{ij} x_i}{\left(\frac{\partial f}{\partial x_i}\right)}.
\end{equation*}
For the analysis, we also increase the dual variable $y_j$
at the rate $\frac{d y_j}{dt} = 1$.

\subsection{Analysis of the Algorithm} 

The first observation follows from our choice of $\gamma$.
\begin{observation}
\label{obv:initialization}
	The value of the objective $f(\vec{x})$ after the initialization is at most $\opt$.
\end{observation}

Our main goal is to bound the total increase of the objective
over the course of the online algorithm. Recall the KKT conditions for 
optimality of convex programs: 
\begin{enumerate}
	\item \label{kkt-feasible} {\bf Feasibility:} $C \vec{x} \geq \vec{1}$, $\vec{x} \geq \vec{0}$, and $\vec{y} \geq \vec{0}$;
	\item \label{kkt-slackness} {\bf Complementary Slackness:} $y_j \cdot \left(\sum_{i=1}^m c_{ij} x_i - 1\right) = 0$ for all $j\in [n]$;
	\item \label{kkt-stationary} {\bf Stationarity:} $\sum_{j=1}^n c_{ij} y_j = \frac{\partial f}{\partial x_i}$ for all $i\in [m]$.
\end{enumerate}
Clearly, the online algorithm maintains feasibility (condition~\ref{kkt-feasible}). 
It will be useful to establish approximate
stationarity (condition \ref{kkt-stationary}) at the end of the algorithm. 

\begin{lemma}
\label{lma:stationary}
	Let $\alpha = \ln (\gamma / c_{\min})$ where $c_{\min} = \min_{i, j} \{c_{ij} > 0\}$,
	and let $\left(\frac{\partial f}{\partial x_i}\right)_e$
	be the value of $\left(\frac{\partial f}{\partial x_i}\right)$ at the end of the algorithm.
	The following holds for all $i\in [m]$:
	\begin{equation}
	\label{eq:stationary-approx}
		\sum_{j=1}^n c_{ij} y_j \leq \alpha \cdot \left(\frac{\partial f}{\partial x_i}\right)_e.
	\end{equation}	
\end{lemma}
\begin{proof}
	Suppose the algorithm is updating variables for constraint $j$ at time $t$. We bound the 
	rate of increase of the LHS of (\ref{eq:stationary-approx}):
	\begin{equation*}
		\frac{d \sum_{j=1}^n c_{ij} y_j}{dt}
		 = \frac{d c_{ij} y_j}{dt} 
		 = c_{ij} 
		= \left( \frac{\partial f}{\partial x_i}\right)_t \cdot (1/x_i) \cdot \frac{d x_i}{dt} 
		\leq \left( \frac{\partial f}{\partial x_i} \right)_e \cdot (1/x_i) \cdot \frac{d x_i}{dt}.
	\end{equation*}
	The last step uses the convexity of $f$, which implies non-decreasing partial derivatives.
	Since the maximum value of any variable $x_i$ can be $1 / c_{\min}$, it follows that
	\begin{equation*}
		\sum_{j=1}^n c_{ij} y_j \leq \left( \frac{\partial f}{\partial x_i} \right)_e \int_{1/\gamma}^{1/c_{\min}} \frac{d x_i}{x_i}.\qedhere
	\end{equation*}
\end{proof}

We start the analysis by comparing the Lagrangian dual to the primal objective.
\begin{lemma}
\label{lma:rate}
	At any stage of the online algorithm,
	\begin{equation}
	\label{eq:rate}
		f(\vec{x}) - f(\vec{x^0})\leq \sum_{j=1}^n y_j.
	\end{equation}
\end{lemma}
\begin{proof}
	We compare the rates of increase of the two sides
	of Eqn.~\ref{eq:rate} in the online algorithm:
	\begin{equation*}
		\frac{d f(\vec{x})}{dt}
		= \sum_{i=1}^m \left(\frac{\partial f}{\partial x_i}\right)_t \cdot \frac{d x_i}{dt}
		= \sum_{i=1}^m \left(\frac{\partial f}{\partial x_i}\right)_t \cdot \frac{c_{ij} x_i}{\left(\frac{\partial f}{\partial x_i}\right)_t} 
		= \sum_{i=1}^m c_{ij} x_i
		\leq 1
		= \frac{d y_j}{dt}.\qedhere
	\end{equation*}
\end{proof}
We are now ready to prove our main lemma.
%
%
\begin{lemma}
\label{lma:approx}
	If $\vec{x^*}$ be any feasible solution and $\vec{x}$ is the solution obtained by the online
	algorithm, then
	\begin{equation*}
	f(\vec{x}) \leq f(\alpha \beta \vec{x^*}) + \beta f(\vec{x^0}), 
	\quad \text{where } \beta = \max_{\vec{x}} \frac{\sum_{i=1}^m x_i \cdot \frac{\partial f}{\partial x_i}}{f(\vec{x})}.
	\end{equation*}
\end{lemma}
\begin{proof}
	By first order convexity propeties,
	\begin{equation*}
		f(\alpha \beta \vec{x^*}) - f(\vec{x})
		\geq \sum_{i=1}^m (\alpha \beta x^*_i - x_i) \frac{\partial f}{\partial x_i} .
	\end{equation*}
	The RHS above can be written as
%
	\begin{equation*}
		\beta \sum_{i=1}^m \left(\alpha x^*_i \frac{\partial f}{\partial x_i} - (x_i / \beta) \frac{\partial f}{\partial x_i}\right) 
		\geq \beta \sum_{i=1}^m \left(x^*_i \sum_{j=1}^n c_{ij} y_j - (x_i / \beta) \frac{\partial f}{\partial x_i}\right) 
		\quad \text {(by Lemma~\ref{lma:stationary})}.
	\end{equation*}
	Swapping summations, the RHS above can be written as
	\begin{equation*}
		\beta \left(\sum_{j=1}^n y_j \sum_{i=1}^m c_{ij} x^*_i  - (1 / \beta)\sum_{i=1}^m x_i \frac{\partial f}{\partial x_i}\right) 
		\geq \beta \left(\sum_{j=1}^n y_j  - (1 / \beta)\sum_{i=1}^m x_i \frac{\partial f}{\partial x_i}\right) 
		\quad (\text {by feasibility of } \vec{x^*}).
	\end{equation*}		
	Using the definition of $\beta$, the RHS above can be written as
	\begin{equation*}
		\beta \left(\sum_{j=1}^n y_j  - f(\vec{x})\right) 
		\geq -\beta f(\vec{x^0})
		\quad \text{(by Lemma~\ref{lma:rate})}.\qedhere
	\end{equation*}
\end{proof}
\noindent
Finally, Theorem~\ref{thm:mainocg} follows from Lemma~\ref{lma:approx} and Observation~\ref{obv:initialization}.

\section{Fractional Algorithm for \gma}
\label{sec:fractional}


Recall that the input contains the pair of values $({\bf C}, {\bf L})$
with the guarantee that there exists a feasible assignment of cost
at most $\bf C$ and $\ell_p$-norm at most $\bf L$. We will fix
such an assignment and call it the {\em optimal} solution
(denoted $\opt$). We
will also assume that the algorithm knows the number of jobs $n$,
which is without loss of generality up to constant factors in the
competitive ratio.

The algorithm has two phases --- an offline pre-processing phase, and
an online phase that (fractionally) schedules the arriving jobs.

\medskip
\noindent
{\bf Offline Pre-processing.}
First, we note that all machines whose startup cost exceeds $\bf C$
are unused in $\opt$; hence, the algorithm discards these machines at
the outset. Let $m$ be redefined to the number of machines with
startup cost at most ${\bf C}$. Next, we multiply the costs of all
machines by $\frac{m}{\bf C}$ so that the cost of $\opt$ is $m$.
%
%
For any machine $i$ with
$c_i \leq 1$, we set $c_i = 1$; this increases the optimal cost to
at most $2m$. We initialize $x_i$ as follows:
if $c_i = 1$, we set $x_i = 1$; else ($1 < c_i \leq m$), we set $x_i = 1/m$.
Finally, we multiply all processing times by $\frac{\beta^{1/p}}{\bf L}$,
where $\beta = \frac{m\ln m}{(40p)^p}$; then an $\ell_p^p$-norm of $\beta$
with the scaled processing times implies an $\ell_p$-norm of $\bf L$
with the original processing times.

Before describing the online phase, we need to introduce some notation.
Let machine $i$ be said to be {\em closed}, {\em partially open}, or
{\em fully open} depending on whether $x_i = 0$, $0 < x_i < 1$ or $x_i = 1$ respectively.
We distinguish between (fractions of) jobs that are assigned when a machine is partially open
and those that are assigned when the machine is fully open;
let us denote the respective sets of jobs $J_0^{(i)}$ and $J_1^{(i)}$. (There can be at most one job that is in
both sets since machine $i$ became fully open while the job was being assigned. For this job, we will consider
the fraction of the job assigned while machine $i$ was partially open as being in set $J_0^{(i)}$ and the
remainder in set $J_1^{(i)}$). Recall that the load on machine $i$ is
$L_i = \sum_{j\in J} y_{ij}p_{ij}$. However, for partially open machines, calculating this
load exactly turns out to be difficult. Instead, we maintain an upper bound of
$c_i^{1/p}x_i$ on the load, which then allows us to define a proxy load
$\simload_i = c_i^{1/p}x_i + \sum_{j \in J_1^{(i)}} y_{ij} p_{ij}$.
%

Suppose the algorithm wants to assign an infinitesimal fraction of a job
to the machines. Intuitively, it should prefer machines whose cost and
fractional $\ell_p$-norm increases the least on assigning the fractional job.
To formalize this notion, we define a function $\psi_{ij}$ that
the algorithm uses to sort machines in increasing order of preference
when assigning a fraction of job $j$:
\begin{equation*}
\psi_{ij} =
\left\{ \begin{array}{ll}
 \max \{c_i^{(p-1)/p} p_{ij}, p_{ij}^p\} & {\rm if}~x_i< 1. \\
    (\simload_i +p_{ij})^p - {\simload_i }^p & {\rm if}~x_i \geq 1.
\end{array}\right.
\end{equation*}

\medskip
\noindent
{\bf Online Assignment.}
When a new job $j$ arrives, we use Algorithm~\ref{proc:frac}
to update $x_i, y_{ij}$ in multiple steps
until $\sum_{i\in M} y_{ij} = 1$.
This is a polynomial-time
implementation of a continuous multiplicative weight augmentation
algorithm, $N$ being the discretization parameter that we set
to $n m \ln m$ to ensure that each discrete step is small enough.
(For technical reasons, we maintain $y_{ij} \leq 2 x_i$ instead of 
$y_{ij} \leq x_i$.)
\renewcommand\footnoterule{} 
\begin{algorithm}[H]

while $\sum_{i\in M} y_{ij} < 1$, do the following:
\begin{itemize}
\item Sort the machines in non-increasing order by $\psi_{ij}$ and
      let $P(j)$ be the minimal prefix\footnotemark[1] of this sorted order such that $\sum_{i\in P(j)} x_i \geq 1$.
\item For each partially\footnotemark[2] open machine $i \in P(j)$, set $\Delta x_i = \frac{x_i}{c_i N}$.

\item For each machine $i \in P(j)$, set $\Delta y_{ij} = \min \left(\frac{x_i}{\psi_{ij} N}, 2 x_i - y_{ij}\right)$.

\item Update $x_i \leftarrow x_i + \Delta x_i $, $y_{ij} \leftarrow y_{ij} + \Delta y_{ij}$, unless $x_i$ or $y_{ij}$ 
exceeds 1. In this case, we do a {\em small step}, i.e., 
we redefine $\Delta x_i$ and $\Delta y_{ij}$ with a value of $N' > N$ instead of $N$ so that 
$\displaystyle\max_{i,j} \{x_i,y_{ij}\} = 1$.

\end{itemize}

\caption{Fractional assignment for a single job}\label{proc:frac}
\end{algorithm}
\footnotetext[1]{$P(j)$ is always defined since $\sum_{i\in M} x_{i} \geq 1$.}
\footnotetext[2]{Only the last machine in $P(j)$ may be fully open; all other machines are partially open.}

%

\subsection{Analysis of the fractional algorithm}

We bound the cost and $\ell_p$-norm of the fractional algorithm using a potential function defined as
$$ \Phi_i = \left\{ \begin{array}{ll}
 c_i x_i & {\rm if}~x_i< 1. \\
 {\simload}_i^p + \sum_{j \in J_1^{(i)}} y_{ij} p_{ij}^p & {\rm if}~x_i = 1.
\end{array}\right. $$
The overall potential function $\Phi = \sum_{ i \in M} \Phi_i$.
Note that the potential function is continuous and monotonically non-decreasing.
First, observe that the potential of a partially open machine is exactly
its fractional startup cost and becomes $c_i$ when the machine is fully opened
(i.e., when $x_i$ becomes 1). Therefore, by monotonicity, $\Phi_i \geq c_i x_i$
during the entire run of the algorithm.
Additionally, the algorithm ensures for each partially open machine, the following
conditions are satisfied:
\begin{equation}
\label{eq:xi-defn-2}
\sum_{j \in J_0^{(i)}} y_{ij} p_{ij} \leq c_i^{1/p} x_i
\quad \text{~and~} \quad
\sum_{j \in J_0^{(i)}} y_{ij} p_{ij}^p \leq c_i x_i.
\end{equation}
Therefore, the potential also bounds the fractional objective function, i.e. the
fractional $\ell_p^p$-norm of the load.

Note that $\psi_{ij}$ is a bound on the discrete differential $\frac{\Delta \Phi_i}{\Delta y_{ij}}$.
For partially open machines,
$\frac{\Delta \Phi_i}{\Delta y_{ij}} = \frac{c_i \Delta x_i}{\Delta y_{ij}}$,
and from the two conditions in Eqn.~\ref{eq:xi-defn-2} we get
$\Delta y_{ij} p_{ij} \leq c_i^{1/p} \Delta x_i$, and $\Delta y_{ij} p_{ij}^p \leq c_i \Delta x_i$,
which defines $\psi_{ij}$.
For fully open machines, the discrete differential is immediate.

First, we bound the increase in potential in the pre-processing phase 
(Lemma~\ref{lma:pot-initial}), in each single step 
step (Lemma~\ref{lma:potential-single}),
and in all the {\em small steps} (Lemma~\ref{lma:smallsteps}).
\begin{lemma}
\label{lma:pot-initial}
At the end of the pre-processing phase, $\Phi \leq m$.
\end{lemma}
\begin{proof}
After pre-processing, the potential $\Phi = \sum_{i\in M} c_i x_i$,
where each $c_i x_i \leq 1$.
\end{proof}

%
\begin{lemma}\label{lma:potential-single}
	The increase in the potential in a single algorithmic step is at most $5/N$.
\end{lemma}
\begin{proof}
The total increase in $\Phi$ for partially open machines in each step is
$$\sum_{i \in PA} c_i \Delta x_i \leq \sum_{i \in P(j)} c_i \frac {x_i}{c_i N}  \leq \frac{2}{N}.$$
For a fully open machine $i$,
$ \Delta y_{ij} = \frac{1}{N(({\simload}_i+p_{ij})^p - {{\simload}_i}^p)} \leq \frac{1}{pNp_{ij}{{\simload}_i}^{p-1}}$,
which increases the first term in $\Phi$ by
\begin{eqnarray*}
({\simload}_{i}+\Delta y_{ij} p_{ij})^p - {{\simload}_i}^p
& \leq & \left({\simload}_i + \frac{1}{Np{{\simload}_i}^{p-1}}\right)^p-{{\simload}_i}^p \\
= {{\simload}_i}^p \left(\left(1+\frac{1}{Np{{\simload}_i}^p}\right)^p - 1\right)
& \leq & {{\simload}_i}^p \left(\left(1+\frac{2}{N{{\simload}_i}^p}\right) - 1\right)
\leq \frac{2}{N}.
\end{eqnarray*}
%
The penultimate inequality follows from $(1+\alpha)^p \leq 1 + 2\alpha p$ for $\alpha \leq 1/(2p)$,
which in turn holds since ${{\simload}_i}^p \geq c_i \geq 1$ and $N \geq 2$.
Additionally, note that
$$\Delta y_{ij} = \frac{1}{N(({\simload}_i+p_{ij})^p - {{\simload}_i}^p)} \leq \frac{1}{N p_{ij}^{p}}.$$
So the increase in the second term of $\Phi_i$ is at most $1/N$.
Further, in each step, the load on at most one fully open machine increases.
Hence, the total increase in potential is at most $5/N$.
\end{proof}
%
\begin{lemma}
\label{lma:smallsteps}
The total increase in potential in all the small steps is at most 2.
\end{lemma}
\begin{proof}
In each small step, either machine becomes fully open or a job is completely assigned.
So, the total number of small steps is at most $n + m$. Therefore, by Lemma~\ref{lma:potential-single},
the total increase in potential in the small steps is at most $\frac{m+n}{N}  < \frac{n+m}{nm} \leq 2$.
\end{proof}

This leaves us with the task of bounding the total number of 
regular (i.e., not small) steps. We classify these steps according 
to an optimal solution (denoted $\opt$). Let $M_{\opt}$ denote the set
of open machines in $\opt$ and $\opt(j)\in M_{\opt}$ be the machine
where job $j$ is assigned to by $\opt$. The three categories are:
\begin{enumerate}
\item $\opt (j) \in P(j)$ and $\opt (j)$ is partially open
\item $\opt (j) \notin P(j)$ and $\opt (j)$ is partially open
\item $\opt (j)$ is fully open
\end{enumerate}

We bound the total increase in potential in each of the
three categories separately.
\begin{lemma}
\label{lma:pot-category1}
The total increase in potential in the first category steps is $O(m \log m)$.
\end{lemma}
\begin{proof}
In any step of the first category, the value of $x_{\opt (j)}$ increases to
$x_{\opt (j)}\left(1 + \frac{1}{c_{\opt (j)} N}\right)$.
Since $x_i$ is initialized to at least $1/m$ for every machine $i$
in the pre-processing phase and $x_i$ cannot exceed 1,
it follows that the total number of steps in the first category is at most
\begin{equation*}
	\sum_{i\in M_{\opt}} c_i N \log m  = O(N m \log m).
\end{equation*}
Using Lemma~\ref{lma:potential-single}, we
conclude that the increase in potential in these steps is $O(m \log m)$.
\end{proof}

\begin{lemma}
\label{lma:pot-category2}
The total increase in potential in the second category steps is $O(m \log m)$.
\end{lemma}
\begin{proof}
For any step, let $Q(j)$ denote the set of machines in $P(j)$
for which $ \Delta y_{ij} = 2 x_i - y_{ij}$, and let $R(j) = P(j) \setminus Q(j) $.
Note that for any job $j$, an algorithmic step for which
$\sum_{i\in Q(j)} x_i \geq 1/2$ must be its last step. This follows from the observation that in this step,
$\sum_{i \in M} (y_{ij} + \Delta y_{ij}) \geq \sum_{i \in Q(j)} (y_{ij} + \Delta y_{ij}) = \sum_{i \in Q(j)} 2x_i \geq 1$.
So, there are at most $n$ steps of this kind.

Now, we bound the number of algorithmic steps where $\sum_{i\in R(j)} x_i \geq 1/2$.
In any such step, using the fact that $\psi_{ij} \leq \psi_{\opt(j)j}$ (otherwise $\opt(j)\in P(j)$),
we have
\begin{equation*}
\sum_{i\in M} \Delta y_{ij}
\geq \sum_{i\in R(j)} \Delta y_{ij}
= \sum_{i\in R(j)} \frac{x_i}{N \psi_{ij} }
\geq \frac{1}{2 N \psi_{\opt(j)j}}.
\end{equation*}
Since $\opt(j)$ is partially open,
$\psi_{\opt(j)j} = \max\{ c^{(p-1)/p}_{\opt(j)}p_{\opt(j)j}, p_{\opt(j)j}^p \}$.
Let $L^{\opt}_i$ be the load on machine $i$ in $\opt$. Summing over all jobs, we have
\begin{align*}
& \sum_{j\in J} \psi_{\opt(j)j}
\leq \sum_{j\in J} \left({c^{(p-1)/p}_{\opt(j)}p_{\opt(j)j}} +p_{\opt(j)j}^p \right)
\leq \sum_{j\in J} {c^{(p-1)/p}_{\opt(j)}p_{\opt(j)j}} + \beta \\
&= \sum_{i\in M} \sum_{j:\opt(j)=i} c^{(p-1)/p}_{i}p_{ij} + \beta 
= \sum_{i\in M} c^{(p-1)/p}_{i} L^{\opt}_i + \beta
\leq m^{(p-1)/p}\beta^{\frac{1}{p}} + \beta \\
&\leq m^{(p-1)/p}\left(\frac{ m\log m}{(40p)^p}\right)^{1/p} + \beta
\leq \frac{m \log^{1/p}{m}}{40p} + \beta
\leq 2 m \log{m},
\end{align*}
where we use H\"{o}lder's inequality (se e.g., \cite{wiki:HoldersInequality})
in the first inequality on the second line.
Therefore, the total number of steps in this category is bounded by
$O(N m \log m)$. By  Lemma~\ref{lma:potential-single},
the total increase in potential in these steps is $O(m \log m)$.
\end{proof}

\begin{lemma}
\label{lma:pot-category3}
The total increase in potential in the third category steps is $O(m \log m)$.
\end{lemma}
\begin{proof}
Define $L^*_i$ as ${\simload}_i$ at the end of the run of the algorithm.
For each fully open machine $i$ define $\psi^*_{ij} = (L^*_i + p_{ij})^p - {L^*_i}^p$.
By convexity of $x^p$, we have $\psi_{ij} \leq \psi^*_{ij}$.
Recall the proof of Lemma~\ref{lma:pot-category2} and the definition $R(j)$.
An identical argument shows that for each step in the third category we have,
$$
\sum_{i\in M} \Delta y_{ij}
\geq \sum_{i\in R(j)} \Delta y_{ij}
= \sum_{i\in R(j)} \frac{x_i}{N \psi_{ij} }
\geq \frac{1}{2 N \psi_{\opt(j)j}}
\geq \frac{1}{2 N \psi^*_{\opt(j)j}  }.
$$
Therefore, by summing over all jobs, the total number of the third category steps is $\sum_{j\in J} 2 N \psi^*_{\opt(j)j}$.
By Lemma~\ref{lma:potential-single}, the total increase in potential in third category steps is $ 10 \left(\sum_{j\in J} \psi^*_{\opt(j)j}\right)$.
Define $\Delta_o \Phi$ as the increase in potential first and second category steps along with the small steps,
and $\Phi_0$ to be the potential after pre-processing. Also, let $\Delta_{3} \Phi$ be the increase in potential in third category steps
and $M_o$ be the set of machines that are fully opened by the algorithm. Then,
\begin{align*}
& \Delta_3 \Phi
\leq 10 \left( \sum_{j\in J} \psi^*_{\opt(j)j} \right)
\leq \left( 10 \sum_{j\in J} \left((L^*_{\opt(j)} + p_{\opt(j)j})^p - {L^*_{\opt(j)}}^p\right) \right)\\
&\leq 10 \left( \sum_{i\in M_{\opt} \cap M_o} \sum_{j:\opt(j)=i} \left( (L^*_i+p_{ij})^p - {L^*_{i}}^p \right) \right)
\leq 10\left( \sum_{i\in M_{\opt} \cap M_o} \left((L^*_i+L^{\opt}_i)^p - {L^*_{i}}^p\right)\right).
\end{align*}
Rearranging the terms,
\begin{align*}
& \left( \frac{\Delta_3 \Phi}{10} + \sum_{i\in M_{\opt} \cap M_o}(L^*_{i})^p \right)^{1/p}
\leq \left(\sum_{i\in M_{\opt} \cap M_o} (L^*_i+L^{\opt}_i)^p \right)^{1/p}
\leq \left(\sum_{i\in M_{\opt} \cap M_o} (L^*_i+L^{\opt}_i)^p \right)^{1/p} \\
&\leq \left(\sum_{i\in M_{\opt} \cap M_o} {L^*_i}^p\right)^{1/p} +\left(\sum_{i\in M_{\opt} \cap M_o} (L^{\opt}_i)^p\right)^{1/p}
\leq \left(\sum_{i\in M_{\opt} \cap M_o} {L^*_i}^p\right)^{1/p} +\beta^{1/p}.
\end{align*}
Now, we have two cases. 
First, suppose  $2 \left(\Delta_3 \Phi \right) > \sum_{i\in M_{\opt} \cap M_o} {L^*_i}^p$. Then, we have
\begin{align*}
& \left(\frac{\Delta_3 \Phi}{10} + \sum_{i\in M_{\opt} \cap M_o}(L^*_{i})^p\right)^{1/p} - \left(\sum_{i\in M_{\opt} \cap M_o}(L^*_{i})^p\right)^{1/p} \\
&\geq \left(\frac{\Delta_3 \Phi}{10} + 2 \left(\Delta_3 \Phi\right)\right)^{1/p} - (2\left(\Delta_3 \Phi\right))^{1/p}  
\geq \frac{(2 \left(\Delta_3 \Phi\right))^{1/p}}{40p}.
\end{align*}
The two last equations imply
$\frac{(2 \left(\Delta_3 \Phi\right))^{1/p}}{40p} \leq \beta^{1/p}$, which implies
$2 \left(\Delta_3 \Phi\right) = O(m \log m)$.
Next, consider $2 \left(\Delta_3 \Phi\right) \leq \sum_{i\in M_{\opt} \cap M_o} {L^*_i}^p$. Then, we have
$$2 \left(\Delta_3 \Phi\right) \leq \sum_{i\in M_{\opt} \cap M_o} {L^*_i}^p \leq \Phi_0 + \Delta_o \Phi + \Delta_3 \Phi,$$
which implies that
$\Delta_3 \Phi \leq \Phi_0 + \Delta_o \Phi = O(m\log m)$
by Lemmas~\ref{lma:pot-initial}, \ref{lma:smallsteps}, \ref{lma:pot-category1}, and \ref{lma:pot-category2}.
\end{proof}

The overall bound on the potential now follows
from Lemmas \ref{lma:pot-initial}, \ref{lma:smallsteps}, \ref{lma:pot-category1}, \ref{lma:pot-category2},
and \ref{lma:pot-category3}.
\begin{theorem}\label{thm:fractional-deltaphi}
At the end of the algorithm, the potential is $O(m \log m) = O((40p)^p\beta)$.
\end{theorem}

\medskip
\noindent
{\bf Bounding the cost and objective function.}
Having provided a bound on the potential function, we now relate it to the fractional cost and
$\ell_p$-norm of machine loads using Lemma~\ref{lma:loadinPA} 
and Lemma~\ref{lma:loadPinPA} respectively.
\begin{lemma}
\label{lma:loadinPA}
For each partially open machine $i$, $\Delta y_{ij} p_{ij}  \leq \Delta x_i {c_i}^{1/p}$.
\end{lemma}
\begin{proof}
In each update step of a partially open machine $i$,
\begin{equation*}
\Delta y_{ij} p_{ij} = \frac{p_{ij} x_i}{\psi_{ij} N} =
\frac{p_{ij} \Delta x_i c_i}{\psi_{ij}} \leq
\frac{p_{ij} \Delta x_i c_i}{c_i^{(p-1)/p} p_{ij}} =
\Delta x_i {c_i}^{1/p}.\qed
\end{equation*}
\end{proof}

\begin{lemma}
\label{lma:loadPinPA}
For each partially open machine $i$, $\Delta y_{ij} p_{ij}^p  \leq \Delta x_i c_i$.
\end{lemma}
\begin{proof}
In each update step of a partially open machine $i$,
\begin{equation*}
\Delta y_{ij} p_{ij}^p = \frac{p_{ij}^p x_i}{\psi_{ij} N} =
\frac{p_{ij}^p \Delta x_i c_i}{\psi_{ij}} \leq
\frac{p_{ij}^p \Delta x_i c_i}{p_{ij}^p} =
\Delta x_i c_i.\qedhere
\end{equation*}
\end{proof}

Finally, we give the overall bound for the fractional 
solution.
\begin{theorem}
\label{thm:fracobjective}
For the fractional solution,
the objective (fractional $\ell_p^p$-norm of loads)
is bounded by $O((40p)^p\beta)$
and the total cost is bounded by $O(m\log m)$.
\end{theorem}
\begin{proof}
The first term in the fractional objective is bounded using Lemma~\ref{lma:loadinPA}. If $x_i < 1$, then
$$\sumds_j \left(\frac{y_{ij} p_{ij}}{x_i}\right)^p x_i \leq \left(\frac{c_i^{1/p} x_i}{x_i}\right)^p x_i = c_i x_i =  \Phi_i.$$
On the other hand, if $x_i=1$, then
$$\sumds_j \left(\frac{y_{ij} p_{ij}}{x_i}\right)^p x_i \leq (c_i^{1/p} + \sum_{j \in J_1^{(i)}} y_{ij}p_{ij})^p = \simload_i^p  \leq \Phi_i.$$
The second term in the fractional objective  is bounded using Lemma~\ref{lma:loadPinPA}:
$$\sumds_j y_{ij} p_{ij}^p = \sumds_{j \in J_0^{(i)}} y_{ij} p_{ij}^p + \sumds_{j \in J_1^{(i)}} y_{ij} p_{ij}^p \leq c_i x_i + \sumds_{j \in J_1^{(i)}} y_{ij} p_{ij}^p \leq \Phi_i.$$
Summing over all machines, the fractional objective is at most $2\Phi= O((40p)^p \beta)$.
Since for all machines, $c_i x_i \leq \Phi_i$,
the total cost is also bounded by the potential function, which is $O(m\log m)$ by Theorem~\ref{thm:fractional-deltaphi}.
\end{proof}



\section{Online Rounding for \gma with $\ell_p$ norm}
\label{sec:rounding}

\begin{algorithm}[t]
\textbf{Opening Machines:}
	For every machine $i$ whose blue copy is closed, open it with probability (w/p)
	$\min\left(\frac{\alpha(x_i(j) - x_i(j-1))}{1 - \alpha \cdot x_i(j-1)}, 1\right)$.
	(Eqn.~\ref{eq:integer-open} is satisfied by this rule using conditional probabilities.)\\
\textbf{Assigning Job $j$:}\\
	- if $\sum_{i\in M^{(1)}(j)} y_{ij} \geq \frac{1}{2}$, then
    assign to blue copy of $i\in M^{(1)}(j)$
    w/p $\frac{y_{ij}}{\sum_{i\in M^{(1)}(j)} y_{ij}}$, \\
    - else if $\sum_{i\in M_o^{(0)}(j)} z_{ij} \geq 1$, then assign
    to blue copy of $i\in M_o^{(0)}(j)$ w/p
	$\frac{z_{ij}}{\sum_{i\in M_o^{(0)}(j)} z_{ij}}$, \\
	- else assign to red copy of
	$i^* = \arg\min_{i\in M} \left(\left(\hat{L}_i + p_{ij}\right)^p - \hat{L}_i^p\right)$
	after opening it.
\caption{Assignment of a Single Job by the Integer Algorithm}
\label{proc:int}
\end{algorithm}

%
%
There are two decisions that an integer algorithm must make on receiving a new job $j$.
First, it needs to decide the set of machines that it needs to open. Note that since
decisions are irrevocable in the online model, the open machines form a monotonically
growing set over time. Next, the algorithm must decide which among the open machines should
it assign job $j$ to. As we describe below, both these decisions are made by the integer
algorithm based on the fractional solution that it maintains using the algorithm given in
the previous section.
Following nomenclature established by Alon~{\em et al}~\cite{AlonAABN09},
we call this process of producing an integer solution online based on a monotonically
evolving fractional solution an {\em online randomized rounding} procedure.

To simplify the analysis later, we will consider two copies of each machine:
a {\em blue} copy and a {\em red} copy. Note that
this is without loss of generality, up to a constant factor loss in the competitive
ratio for both the cost and $\ell_p$-norm objectives. First, we define a randomized
process that controls the opening of blue copies of machines in the integer algorithm.
Let $M_o(j)$ denote the set of machines whose blue copies are open after job $j$
has been assigned, and $X_i(j)$ be an indicator random variable
whose value is 1 if machine $i\in M_o(j)$ and 0 otherwise. Let $x_i(j)$ be the value of
variable $x_i$ in the fractional solution after job $j$ has been completely
assigned (fractionally). The integer algorithm maintains the invariant
\begin{equation}
\label{eq:integer-open}
	\pr[X_i(j) = 1] =  \min(\alpha \cdot x_i(j), 1)
	\text{~for some parameter~} \alpha \text{~that we will set later}.
\end{equation}
using the rule given in Algorithm~\ref{proc:int}. Next,
we need to assign job $j$ to one of the open machines. We partition the set of
machines $M$ into two sets based on the fractional solution:
$M^{(0)}(j)$ represents machines $i$ such that $x_i(j) < \frac{1}{\alpha}$
and $M^{(1)}(j)$ represents machines $i$ such that
$x_i(j) \geq \frac{1}{\alpha}$. Note that after job $j$, the blue copies of all
machines in $M^{(1)}(j)$ are open (by Eqn.~\ref{eq:integer-open}).
On the other hand, the blue copies of
some subset of machines in $M^{(0)}(j)$ are open; call this subset $M_o^{(0)}(j)$,
i.e. $M^{(0)}(j) = M^{(0)}(j) \cap M_o(j)$.
In addition let, $z_{ij} = \frac{4 y_{ij}}{\alpha \cdot x_i}$ and
$\hat{L}_i$ be the current sum of processing times of all jobs assigned to the red copy of machine $i$.
The assignment rule for job $j$ is given in Algorithm~\ref{proc:int}. 

\subsection{Analysis}
First, we argue about the expected cost of the solution.
To bound the cost of red copies, we show that Case 3 has low probability.
\begin{lemma}
\label{lma:case3-prob}
	For any job $j$, the probability of case 3 is at most $\exp(-\alpha/48)$.
\end{lemma}
\begin{proof}
	Consider a machine $i\in M^{(0)}(j)$,
	i.e. $x_i(j) < \frac{1}{\alpha}$. Such a machine is open after
	job $j$ with probability $\alpha x_i(j)$. Let us define a corresponding
	random variable
	\begin{equation*}
			Z_{ij} = \left\{\begin{array}{cc}
				z_{ij} &  \text{if~} i\in M_o^{(0)}(j) \\
				0 & \text{otherwise.}
			\end{array}\right.
	\end{equation*}
	We need to bound the probability that $\sum_{i\in M^{(0)}(j)} Z_{ij} < 1$.
	
	First, we observe that
	$Z_{ij} \leq \frac{4 y_{ij}}{\alpha \cdot x_i(j)} \leq \frac{8}{\alpha}$
	since $y_{ij} \leq x_i(j)$. Now, consider random variable
	\begin{equation*}
			\tilde{Z}_{ij} = \left\{\begin{array}{cc}
				\frac{8}{\alpha} &  \text{with probability~} \frac{\alpha y_{ij}}{2} \\
				0 & \text{otherwise.}
			\end{array}\right.
	\end{equation*}
	Note that the expectations of $Z_{ij}$ and $\tilde{Z}_{ij}$ are identical
	and both have only one non-zero value in their support,
	but $Z_{ij}$ has a strictly smaller range. Therefore, any tail bounds that
	apply to $\tilde{Z}_{ij}$ also apply to $Z_{ij}$. Further, note that
	\begin{equation*}
		\ex\left[\sum_{{i\in M^{(0)}(j)}} Z_{ij}\right]
		= \ex\left[\sum_{{i\in M^{(0)}(j)}} \tilde{Z}_{ij}\right]
		= 4 \sum_{{i\in M^{(0)}(j)}} y_{ij}
		\geq 2.
	\end{equation*}
	Therefore, by Chernoff-Hoeffding bounds (e.g.,~\cite{MotwaniR97}),
	\begin{align*}
		& \pr\left[ \sum_{{i\in M^{(0)}(j)}} Z_{ij} < 1 \right]
		\leq \pr\left[ \sum_{{i\in M^{(0)}(j)}} \tilde{Z}_{ij} < 1 \right]
		= \pr\left[ \sum_{{i\in M^{(0)}(j)}} \frac{\alpha}{8} \tilde{Z}_{ij} < \frac{\alpha}{8} \right] \\
		&\leq \exp\left(-\frac{\frac{1}{2^2}\cdot \frac{2 \cdot \alpha}{8}}{3}\right)
		= \exp\left(-\alpha/48\right).
	\end{align*}
\end{proof}
We choose $\alpha = 48 \ln (mn)$ to obtain the following corollary.
(For now, $\alpha = 48 \ln n$ would have sufficed but we will need
$\alpha \geq \ln m$ in a later step.)
\begin{corollary}
\label{cor:case3-prob}
	For any job $j$, the probability of case 3 is at most $\frac{1}{mn}$.
\end{corollary}
Recall that the cost of each individual machine is at most $m$.
Using linearity of expectation and the above corollary, we can now claim
that the expected cost of red copies of machines is at most $m$.
Similarly, using linearity of expectation and Eqn.~\ref{eq:integer-open},
we can claim that the expected cost of blue copies of machines is
$\sum_{i\in M} c_i \alpha x_i \leq \alpha \Phi$.
Overall, we get the following bound for
the cost of machines opened by the integer algorithm.
\begin{lemma}
\label{lma:cost-integer}
	The total expected cost of machines in the integer algorithm is
	at most $(\alpha + 1) \Phi = O(\ln (mn)) \Phi$.
\end{lemma}

We are now left to bound the expected $\ell_p$-norm of machine loads.
First, we obtain a bound for red copies of machines. Note that the
assignment of jobs in Case 3 follows a greedy algorithm assuming all
machines are open. Therefore, the analysis of the $\ell_p$-norm of
the red machines follows directly from the corresponding analysis
without startup costs~\cite{Caragiannis08}.
\begin{lemma}[\cite{Caragiannis08}]
\label{lma:red-load}
	The competitive ratio of the $\ell_p$-norm of the red copies of
	machines is $O(p)$.
\end{lemma}

Finally, we will bound the $\ell_p$-norm of blue copies of machines.
Let us define an indicator random variable
\begin{equation*}
	Y_{ij} = \left\{\begin{array}{ll}
		1 & \text{if job~} j \text{~is assigned to the blue copy of machine~} i \text{~in the integer solution} \\
		0 & \text{otherwise}.
	\end{array}\right.
\end{equation*}
Then, the $\ell_p^p$-norm of the integer solution can be written as
$\sum_{i\in M} \left(\sum_{j\in J} Y_{ij} p_{ij}\right)^p$.
We will bound the expected $\ell_p^p$-norm of each machine
individually, and then use linearity of expectation over all
the machines.

Our main technical tool in bounding the expected $\ell_p^p$-norm of a single machine
will be the following theorem (see e.g.,~\cite{JohnsonSZ85} for a proof).
\begin{theorem}
\label{thm:lp}
Let $W_1, W_2, \ldots, W_n$ be independent non-negative random variables.
Let $p > 1$ and $K_p = \Theta\left(\frac{p}{\log p}\right)$.
Then,
\begin{equation*}
	\ex\left[\left(\sum_{j=1}^n W_j\right)^p\right]
	\leq \left(K_p\right)^p \cdot \max\left(\left(\sum_{j=1}^n \ex[W_j]\right)^p,
	\sum_{j=1}^n \ex\left[\left(W_j\right)^p\right]\right).
\end{equation*}	
\end{theorem}
Ideally, we would like to use this theorem directly with $W_j = Y_{ij} p_{ij}$.
This is indeed possible if $x_i(j) \geq \frac{1}{\alpha}$ since the assignment
of such jobs $j$ to machine $i$ are independent of each other.
However, the assignment of jobs $j$
for which $x_i(j) < \frac{1}{\alpha}$ are {\em not} independent; they depend
on each other via the random variable $X_i$ which denotes whether machine $i$
is open or not. Let $j_i$ be the job that opened machine $i$, i.e.
$X_i(j_i - 1) = 0$ and $X_i(j_i) = 1$. Conditioned on $j_i$, the variables
$Y_{ij}$ for $j \geq j_i$ are indeed independent. First, we will reduce the
conditioning to a single indicator random variable. Define an
indicator random variable $X_i = 1$ if and only if machine $i$ is open
in the integer solution after all $n$ jobs have been assigned.
By Eqn.~\ref{eq:integer-open},
$X_i = 1$ with probability $\min(\alpha x_i, 1)$, where $x_i$ is the fractional
variable after all $n$ jobs have been (fractionally) assigned.
Now, define a binary random variable $\tilde{Y}_{ij}$ with the following
properties:
\begin{itemize}
	\item if $X_i = 0$, then $\tilde{Y}_{ij} = 0$,
	\item else if job $j$ is not assigned via case 2, then $\tilde{Y}_{ij} = Y_{ij}$,
	\item else, $\tilde{Y}_{ij} = 1$ with probability $z_{ij}$;
	furthermore, in this case, using shared randomness, we ensure that
	$\tilde{Y}_{ij} = 1$ whenever $Y_{ij} = 1$.
\end{itemize}
The last condition can be met since in case 2,
	$\pr[Y_{ij} = 1] = \frac{z_{ij}}{\sum_{i\in M_o^{(0)}(j)} z_{ij}} \leq z_{ij}$.
%
Note that conditioned on $X_i = 1$, $\tilde{Y}_{ij}$ for different jobs $j$ are independent
random variables. Furthermore, $\tilde{Y}_{ij}$ {\em stochastically dominates} $Y_{ij}$,
i.e. $Y_{ij} = 1$ implies $\tilde{Y}_{ij} = 1$. Therefore, it suffices to bound
$\left(\sum_{j\in J} \ex[\tilde{Y}_{ij} p_{ij}]\right)^p$
and $\sum_{j\in J} \ex[\tilde{Y}_{ij}] \left(p_{ij}\right)^p$,
conditioned on $X_i = 1$. In the next lemma, we bound the first term.
\begin{lemma}
\label{lma:conditioned-1}
For any machine $i$, conditioned on the event $X_i = 1$, we have
	$$\left(\sum_{j\in J} \ex[\tilde{Y}_{ij} p_{ij}]\right)^p
	\leq \left(5 c_i^{1/p} + \sum_{j\in J^{(i)}_1} y_{ij} p_{ij}\right)^p.$$
\end{lemma}
\begin{proof}
	We consider two phases for machine $i$: $x_i < 1$ and $x_i = 1$. Recall
	that the jobs assigned in the first phase are denoted $J^{(i)}_0$ and
	those in the second phase are denoted $J^{(i)}_1$. First, we note
	that for jobs $j\in J^{(i)}_1$, $\ex[\tilde{Y}_{ij}] \leq y_{ij}$.
	Therefore,
		$\sum_{j\in J^{(i)}_0} \ex[\tilde{Y}_{ij} p_{ij}]
		\leq \sum_{j\in J^{(i)}_0} y_{ij} p_{ij}$.
	On the other hand, for jobs $j\in J^{(i)}_0$, we need to distinguish
	between jobs assigned via case 2 while $x_i(j) < \frac{1}{\alpha}$
	(call this set $J^{(i)}_0 (2)$)
	and those that are assigned via case 3 after $x_i(j) \geq \frac{1}{\alpha}$
	(call this set $J^{(i)}_0 (3)$). Then,
	\begin{align*}
		& \sum_{j\in J^{(i)}_0} \ex[\tilde{Y}_{ij} p_{ij}]
		\leq \sum_{j\in J^{(i)}_0 (2)} z_{ij} p_{ij} +  \sum_{j\in J^{(i)}_0 (3)} y_{ij} p_{ij}
		\leq \frac{4}{\alpha}\sum_{j\in J^{(i)}_0 (2)} \frac{y_{ij} p_{ij}}{x_i(j)} +  \sum_{j\in J^{(i)}_0 (3)} y_{ij} p_{ij} \\
		&\leq \frac{4}{\alpha} \int_{1/m}^{1/\alpha} c_i^{1/p} \frac{d x}{x} + c_i^{1/p} \left(1 - \frac{1}{\alpha}\right)
		\leq  \frac{4}{\alpha} \int_{1/m}^1 c_i^{1/p} \frac{d x}{x} + c_i^{1/p}
		\leq 5 c_i^{1/p},
	\end{align*}
	since $\alpha = 48 \ln (mn) \geq \ln m$. Combining all jobs,
		$$\left(\sum_{j\in J} \ex[\tilde{Y}_{ij} p_{ij}]\right)^p
		\leq \left(5 c_i^{1/p} + \sum_{j\in J^{(i)}_1} y_{ij} p_{ij}\right)^p.$$
\end{proof}
Next, we bound $\sum_{j\in J} \ex[\tilde{Y}_{ij}] \left(p_{ij}\right)^p$,
conditioned on $X_i = 1$.
\begin{lemma}
\label{lma:conditioned-2}
For any machine $i$, conditioned on the event $X_i = 1$, we have
	$$\sum_{j\in J} \ex[\tilde{Y}_{ij}] \left(p_{ij}\right)^p
	\leq 5c_i + \sum_{j\in J^{(i)}_1} y_{ij} \left(p_{ij}\right)^p.$$
\end{lemma}
\begin{proof}
	As in the previous proof,
	we consider two phases for machine $i$: $x_i < 1$ and $x_i = 1$. As earlier,
	the set of jobs assigned in the first phase is denoted $J^{(i)}_0$ and
	that assigned in the second phase is denoted $J^{(i)}_1$. First, we note
	that for jobs $j\in J^{(i)}_1$, $\ex[\tilde{Y}_{ij}] \leq y_{ij}$.
	Therefore,
		$\sum_{j\in J^{(i)}_0} \ex[\tilde{Y}_{ij} \left(p_{ij}\right)^p]
		\leq \sum_{j\in J^{(i)}_0} y_{ij} \left(p_{ij}\right)^p$.
	On the other hand, for jobs $j\in J^{(i)}_0$, we need to distinguish
	between jobs assigned via case 2 while $x_i(j) < \frac{1}{\alpha}$
	(called $J^{(i)}_0 (2)$)
	and those that are assigned via case 3 after $x_i(j) \geq \frac{1}{\alpha}$
	(called $J^{(i)}_0 (3)$). Then,
	\begin{align*}
		& \sum_{j\in J^{(i)}_0} \ex[\tilde{Y}_{ij}] \left(p_{ij}\right)^p
		\leq \sum_{j\in J^{(i)}_0 (2)} z_{ij} \left(p_{ij}\right)^p +  \sum_{j\in J^{(i)}_0 (3)} y_{ij} \left(p_{ij}\right)^p \\
		&\leq \frac{4}{\alpha}\sum_{j\in J^{(i)}_0 (2)} \frac{y_{ij} \left(p_{ij}\right)^p}{x_i(j)} +  \sum_{j\in J^{(i)}_0 (3)} y_{ij} \left(p_{ij}\right)^p 
		\leq \frac{4}{\alpha} \int_{1/m}^{1/\alpha} c_i \frac{d x}{x} + c_i (1 - 1/\alpha) \\
		&\leq \quad \frac{4}{\alpha} \int_{1/m}^1 c_i \frac{d x}{x} + c_i
		\leq 5 c_i,
	\end{align*}
	since $\alpha = 48 \ln (mn) \geq \ln m$. Combining all jobs,
		$$\sum_{j\in J} \ex[\tilde{Y}_{ij}] \left(p_{ij}\right)^p
		\leq 5 c_i + \sum_{j\in J^{(i)}_1} y_{ij} \left(p_{ij}\right)^p.$$
\end{proof}
Finally, we apply Theorem~\ref{thm:lp} to Lemmas~\ref{lma:conditioned-1}
and \ref{lma:conditioned-2}, and remove the conditioning on $X_i$.
\begin{theorem}
For any machine $i$,
	$$\ex\left[\left(\sum_{j\in J} Y_{ij} p_{ij}\right)^p\right]
	\leq ((5 \alpha)^{1/p} K_p)^p \Phi_i,$$
where $K_p = \theta\left(\frac{p}{\log p}\right)$.
\end{theorem}
\begin{proof}
	For any machine $i$, conditioned on the event $X_i = 1$, we have
	\begin{align*}
		& \ex\left[\left(\sum_{j\in J} Y_{ij} p_{ij}\right)^p\right]
		\leq
		\ex\left[\left(\sum_{j\in J} \tilde{Y}_{ij} p_{ij}\right)^p\right]
		\quad (\text{since~} \tilde{Y}_{ij} \text{~stochastically dominates~} Y_{ij}) \\
		&\leq \left(K_p\right)^p	\max\left(		
		\left(\sum_{j\in J} \ex[\tilde{Y}_{ij} p_{ij}]\right)^p,
		\sum_{j\in J} \ex[\tilde{Y}_{ij}] \left(p_{ij}\right)^p\right)
		\quad (\text{using Theorem~\ref{thm:lp}})	\\
		&\leq \left(K_p\right)^p		
		\max\left(\left(5 c_i^{1/p} + \sum_{j\in J^{(i)}_1} y_{ij} p_{ij}\right)^p,
		5c_i + \sum_{j\in J^{(i)}_1} y_{ij} \left(p_{ij}\right)^p\right).
	\end{align*}	
	We now have three cases. First, suppose machine $i$ satisfies $x_i = 1$ after
	all the jobs have been fractionally assigned. Then $X_i = 1$ deterministically,
	and the above inequality holds unconditionally. Therefore, 
	\begin{align*}
		& \ex\left[\left(\sum_{j\in J} Y_{ij} p_{ij}\right)^p\right]
		\leq \left(K_p\right)^p
		\max\left(\left(5 c_i^{1/p} + \sum_{j\in J^{(i)}_1} y_{ij} p_{ij}\right)^p,
		5c_i + \sum_{j\in J^{(i)}_1} y_{ij} \left(p_{ij}\right)^p\right) \\
		&\leq \left(5 K_p\right)^p \Phi_i.
	\end{align*}
	Next, consider machines $i$ such that $\frac{1}{\alpha} \leq x_i < 1$ after
	all the jobs have been fractionally assigned. As in the previous case,
	$X_i = 1$ deterministically,
	and therefore the above inequality holds unconditionally. However,
	for such machines, $J^{(i)}_1 = \emptyset$. Therefore,
	\begin{align*}
		& \ex\left[\left(\sum_{j\in J} Y_{ij} p_{ij}\right)^p\right]
		\leq \left(K_p\right)^p  \max\left(\left(5 c_i^{1/p}\right)^p, 5c_i\right)	\\
		&= 5 \left(K_p\right)^p  c_i
		\leq \left((5 \alpha)^{1/p} K_p\right)^p \Phi_i.
	\end{align*}
	Finally, consider machines $i$ such that $x_i < \frac{1}{\alpha}$ after
	all the jobs have been fractionally assigned. As in the previous case,
	for such machines, $J^{(i)}_1 = \emptyset$. However, $X_i = 1$ with
	probability $\alpha x_i$. Therefore, 
	\begin{align*}
		& \ex\left[\left(\sum_{j\in J} Y_{ij} p_{ij}\right)^p\right]
		= \ex\left[\left(\sum_{j\in J} Y_{ij} p_{ij}\right)^p \Bigg | X_i = 1\right] \cdot \pr[X_i = 1] \\
		&\leq \left(K_p\right)^p (5 c_i) \alpha x_i
		\leq \left((5 \alpha)^{1/p} K_p\right)^p \Phi_i.
	\end{align*}	
\end{proof}
This completes the proof of Theorem~\ref{thm:main}.

\section{Online Rounding for \gma with $\ell_1$ norm}
\label{sec:roundingp1}
We now present an online rounding algorithm specifically tailored
to the important special case of $p = 1$, i.e., the $\ell_1$-norm.
The rule for opening machines is identical (with a smaller value
of $\alpha$ that we will shortly calculate) to the rounding
algorithm for general $p$. However, the assignment rule for a job
is now simpler and is given in Algorithm~\ref{proc:int-l1}.
Here, $M(j)$ denotes the machines sorted in non-decreasing
order of $p_{ij}$ and $M_{1/2}(j)$ is the minimal prefix
of $M(j)$ that satisfies
$\sum_{i\in M_{1/2}(j)} y_{ij} \geq 1/2$. As earlier,
for clarity, we use two copies of each machine, a blue copy
and a red copy, and let $M_o(j)$ be the machines whose
blue copies are open after job $j$. 
%
%

\begin{algorithm}[t]
\textbf{Opening Machines:}
	For every machine $i$ whose blue copy is closed, open it with probability
	$\min\left(\frac{\alpha(x_i(j) - x_i(j-1))}{1 - \alpha \cdot x_i(j-1)}, 1\right)$.
	(Eqn.~\ref{eq:integer-open} is satisfied by this rule using	conditional probabilities.\\
\textbf{Assigning Job $j$:}\\
	- if $M_o(j) \cap M_{1/2}(j) \not= \emptyset$, then assign to blue
	copy of any machine in $M_o(j) \cap M_{1/2}(j)$,\\
	- else assign to red copy of machine $i^* = \arg\min_{i\in M} p_{ij}$,
	after opening it if necessary.
\caption{Assignment of a Single Job by the Integer Algorithm for the $\ell_1$-norm}
\label{proc:int-l1}
\end{algorithm}

\subsection{Analysis}\label{apx:rounding-l1}
First, we argue about the expected cost of the solution.
To bound the cost of red copies, we show that Case 2 has low probability.
\begin{lemma}
\label{lma:case2-prob-l1}
	For any job $j$, the probability of case 2 is at most $\exp(-\alpha/4)$.
\end{lemma}
\begin{proof}
Note that
$$\sum_{i\in M_{1/2}(j)} x_i(j) \geq \sum_{i\in M_{1/2}(j)} \frac{y_{ij}}{2} \geq \frac{1}{4}.$$
Therefore, the probability of case 2 is
\begin{align*}
	& \prod_{i\in M_{1/2}(j)} (1 - \alpha x_i(j))
	\leq \left(1 - \frac{\alpha \sum_{i\in M_{1/2}(j)} x_i(j)}{k}\right)^k \\
	&\leq \exp\left(-\alpha \sum_{i\in M_{1/2}(j)} x_i(j)\right)
	\leq \exp\left(-\alpha/4\right).
\end{align*}
\end{proof}
We choose $\alpha = 4 \ln n$ to obtain the following corollary.
\begin{corollary}
\label{cor:case2-prob-l1}
	For any job $j$, the probability of case 2 is at most $\frac{1}{n}$.
\end{corollary}
Recall that the cost of each individual machine is at most $m$.
Using linearity of expectation and the above corollary, we can now claim
that the expected cost of red copies of machines is at most $m$.
Similarly, using linearity of expectation and Eqn.~\ref{eq:integer-open},
we can claim that the expected cost of blue copies of machines is
$\sum_{i\in M} c_i \alpha x_i \leq \alpha \Phi$.
Overall, we get the following bound for
the cost of machines opened by the integer algorithm.
\begin{lemma}
\label{lma:cost-integer-l1}
	The total expected cost of machines in the integer algorithm is
	at most $(\alpha + 1) \Phi = O(\ln n) \Phi$.
\end{lemma}

We are now left to bound the $\ell_1$-norm of the assignment.
First, consider the red copies of machines. Note that the
assignment of jobs in Case 2 follows a greedy algorithm assuming all
machines are open. Therefore, the $\ell_1$-norm of red copies
of machines is optimal. The next lemma complements this observation
by bounding the $\ell_1$-norm of blue copies of machines.

\begin{lemma}
\label{lma:load-l1}
	The expected $\ell_1$-norm of blue copies of machines is at most $2 \Phi$.
\end{lemma}
\begin{proof}
Suppose we assigned job $j$ to the blue copy of machine $\hat{i}$.
Also, let $k(j)$ be the last machine in the prefix $M_{1/2}(j)$
and let $\overline{M}_{1/2}(j) = (M \setminus M_{1/2}(j)) \cup \{k(j)\}$.
Then, we have
	$\sum_{i\in \overline{M}_{1/2}(j)} y_{ij} \geq 1/2$ by minimality of the prefix $M_{1/2}(j)$
	and $p_{\hat{i}j} \leq p_{ij}$ for all machines $i\in \overline{M}_{1/2}(j)$.
Then, the increase in $\ell_1$-norm of the integer solution is
is $p_{\hat{i}j}$ whereas the corresponding increase in $\Phi$ for the fractional
solution is
\begin{equation*}
\sum_{i \in M} y_{ij} p_{ij}
\geq \sum_{i \in \overline{M}_{1/2}(j)} y_{ij} p_{ij}
\geq p_{\hat{i}j} \sum_{i \in \overline{M}_{1/2}(j)} y_{ij}
\geq \frac{p_{\hat{i}j}}{2}.
\end{equation*}
The lemma now follows by summing over all jobs.
\end{proof}
This completes the proof of Theorem~\ref{thm:l1}.

\bibliographystyle{plain}
\bibliography{ref}


\appendix
\section*{Appendix}

\section{Lower bound for \ompc with the $\ell_p$ norm objective}
We adapt the example in Azar~{\em et al.} \cite{AzarBFP13} 
for the $\ell_\infty$ norm and analyze it for the $\ell_p$ norm.
For parameters $p$ and $d$, the example uses $r (\geq 2^p)$ packing constraints,
each with at most $\hat{d} = d\log r$ variables and at most $2d$ (which is $< \hat{d}$) variables in any covering constraint.
The example uses $2(r-1)$ pairwise disjoint sets (blocks) of $d$ variables. We use $B_i$ to refer to the $i$th block.
In \cite{AzarBFP13} there is a procedure of revealing covering constraints
to two blocks such that at least one block has a weight of at least $H_d/2$, where $H_d$ refers to the $d$th harmonic number, and there is feasible solution with total weight of $1$ to one of the blocks.

The packing constraints are represented as follows:
a complete binary tree with $r$ leaf nodes. Each node in this tree
except the root corresponds to a block, and no two nodes correspond to the same block. Our packing
constraints correspond to the leaf nodes, with packing constraint $k$ being $\sum(\cup_{i\in Q_k} B_i) \leq \lambda$ where $Q_k$ is the
set of blocks encountered on the path from the root to the leaf node corresponding to packing constraint $k$.
In the example, initially apply the procedure to the two blocks
which are the children of the roots. Then, apply the procedure to the children of the block with the larger weight ($\geq H_d/2$)
 and so on, until it reaches to one of the leafs.
It is easy to verify that for each $1\leq i\leq \log r$ there exists $2^{\log r -1 -i}$ packing constraints with $\lambda \geq i \cdot H_d/2$. In addition, there exists a feasible solution with $\lambda_k = 1$ for any $k$. This yields a competitive ratio of at least (for $p\leq \log r$)
$$\left(\frac{ (H_d/2 \cdot\log r )^p + \displaystyle\sum_{i=1}^{\log r} (H_d/2 \cdot i)^p 2^{\log r -1 -i}}{r}\right)^{1/p} \approx H_d/2 \cdot \frac{p\log e}{e} = \theta(p \log (\hat{d}/\log r)).$$


%
%
%
%
%
%

\end{document}